\documentclass[10pt,journal,compsoc]{IEEEtran}
\ifCLASSOPTIONcompsoc
  \usepackage[nocompress]{cite}
\else
  \usepackage{cite}
\fi
\ifCLASSINFOpdf
   \usepackage[pdftex]{graphicx}
   \graphicspath{{./imgs/}}
   \DeclareGraphicsExtensions{.pdf,.jpeg,.png,.eps}
\else
\fi
\newcommand{\p}[1]{\textsc{p}\oldstylenums{#1}}

\newcommand{\imgid}[1]{\textsf{\textcolor{img_id}{#1}}}

\usepackage{hyperref}
\hypersetup{colorlinks,allcolors=black}

\usepackage{xcolor}
\definecolor{img_id}{RGB}{33,173,90}

\begin{document}
%
\title{Hidden bawls, whispers, and yelps: \\can text be made to sound more \\than just its words?}
%
%
%
%

\author{Caluã~de~Lacerda~Pataca,
        Paula~Dornhofer~Paro~Costa
\IEEEcompsocitemizethanks{\IEEEcompsocthanksitem Caluã de Lacerda Pataca and Paula Dornhofer Paro Costa are with the University of Campinas. They can be contacted at \href{mailto:calua.pataca@gmail.com}{calua.pataca@gmail.com} and \href{mailto:paulad@unicamp.br}{paulad@unicamp.br}.}
}

%
%

\markboth{}%
{}
%



\IEEEtitleabstractindextext{%
\begin{abstract}
Whether a word was bawled, whispered, or yelped, captions will typically represent it in the same way. If they are your only way to access what is being said, subjective nuances expressed in the voice will be lost. Since so much of communication is carried by these nuances, we posit that if captions are to be used as an accurate representation of speech, embedding visual representations of paralinguistic qualities into captions could help readers use them to better understand speech beyond its mere textual content. This paper presents a model for processing vocal prosody (its loudness, pitch, and duration) and mapping it into visual dimensions of typography (respectively, font-weight, baseline shift, and letter-spacing), creating a visual representation of these lost vocal subtleties that can be embedded directly into the typographical form of text. An evaluation was carried out where participants were exposed to this \emph{speech-modulated typography} and asked to match it to its originating audio, presented between similar alternatives. Participants (n=117) were able to correctly identify the original audios with an average accuracy of~65\%, with no significant difference when showing them modulations as animated or static text. Additionally, participants' comments showed their mental models of speech-modulated typography varied widely.
\end{abstract}

}

\maketitle

\IEEEdisplaynontitleabstractindextext

%
\IEEEpeerreviewmaketitle

\IEEEraisesectionheading{\section{Introduction}\label{sec:introduction}}

%
%
%
%
\IEEEPARstart{D}{espite} its expressive richness, when speech is represented through captions it is typically reduced to its words, and its words only. Whatever nuance was originally conveyed by the ways in which the speaker modulated their voice --- their mood, emotions, dispositions, etc --- is lost in this flattened textual representation. This is particularly relevant when captions are used not as a complement to a readily available audio channel, but as its \emph{replacement}. This can be true for deaf and hard of hearing (\textsc{dhh}) persons, but will potentially affect anyone, including hearing individuals facing a situational hearing impairment, e.g., someone affected by a situational hearing impairment such as watching a film on their mobile phone in a noisy environment \cite{biswas2014interface}. If so much of communication is expressed in nuances not captured by written text, what is to be said of the experience of those who have no direct access to acoustic speech but only to its written forms? 

Which is not to say that not capturing \emph{all} of prosody\footnote{Prosody models the paralinguistic dimensions of \emph{how} words are said, beyond their linguistic dimensions of \emph{what} words are said.} in speech is a shortcoming of written text. 
While both speech and text are of course tied together, one is not a simple variant of the other \cite{seidenberg2017}, and just as there are nuances of speech not captured by written text, the opposite is also frequently true, especially now that so much of internet-based communication centers around text \cite{mcculloch2019}. In fact, it has been argued that, while at its origins written text's functions were that of supporting oral speech \cite{nunlist1991}, as the medium developed its uses shifted --- e.g., the invention of modern punctuation and white spaces between words allowed for the emergence of \emph{silent reading} \cite{kuster2016, McCutcheon2015}, a form detached from text's origins in sounded speech.


Still, there are contexts where this gap between speech and what is effectively captured by writing can be an obstacle, as has been shown when \textsc{dhh} individuals point out that captions are not a functional equivalent to hearing since they lack some meta-speech information such as `speaker identification[,] punctuation, sentiment, [and] tone' \cite{kushalnagar2020teleconference}. This is not a new notion (see, for instance, \cite{murphy1983impact}'s speculations from back in 1983 about how an enhanced captioning system could use `special effects, color, and capital letters [to represent] the rich tonal information denied deaf children[.]'), but is becoming more and more relevant when we consider how real-time captioning has grown as a critical assistive technology. As the capabilities of automatic speech recognition (\textsc{asr}) systems increase, captions are finding use in many novel contexts as a way of giving visual access to sounded speech for \textsc{dhh} individuals \cite{loizides2020breaking, mallory2017personal}.

As such, researchers have been exploring ways of enhancing the display of captions to include cues to inform both speaker behavior \cite{seita2018behavioral, mcdonnell2021social} and readers' interpretations. Our work rests on this second case: creating a model capable of representing prosody over regular captions.

We propose a novel model of \emph{Speech-Modulated Typography}, where acoustic features from speech are used to modulate the visual appearance of text. This could allow for a given utterance's transcription to not only represent words being said, but \emph{how} they were said. With this, we hope to uncover typographic parameters that can be generally recognized as visual proxies for the prosodic features of amplitude, pitch, and duration.  Our broad aim is to help further the development of real-time captioning technology, for which the use of algorithmic systems for both processing of prosody and the subsequent modulation of typographic parameters is a requirement. 

This approach can lead to better captioning systems that, just as traditional captions are already useful for \textsc{dhh} \emph{and} hearing individuals \cite{gernsbacher2015video}, have potential benefits for anyone. While we do not aim to represent emotions directly, we argue that, in visually representing an acoustic correlate to emotion, our work is tied to applications and discussions of Affective computing, with additional insights from Visual design and Linguistics. Additionally, and even though our model was developed and evaluated in a Brazilian Portuguese context, our findings are plausibly applicable to any language that uses alphabets.

In this work, we also propose a method to empirically evaluate whether these visual modulations of typography are sufficiently clear to allow for their recognition and understanding by untrained participants. This, we posit, is an important condition if this technology is to eventually gain traction. This methodology informed the creation of a dataset of expressive vocal readings of poetry, which was used in an evaluation conducted with 117 participants aimed at answering two research questions:

\begin{enumerate}
    \item Can readers of this speech-modulated typography use it to recognize prosodic features present in its originating audio but not in its textual content?
    \item Are there differences in recognition between animated and static based versions of speech-modulated typography?
\end{enumerate}


Lastly, in said experiment we also collected open-ended comments from participants about their impressions about the model and its possible uses. These aimed to better qualify the answers we obtained to our research questions, but also to uncover unforeseen potentials and limitations of our model not captured by our quantitative methods.


\subsection{Outline of this paper}

In Section~\ref{sec:related-work} we present an overview of key concepts, challenges, and related work in fields such as linguistics, speech emotion recognition, and design and accessibility. 

Following, in Section~\ref{sec:model} we will explore our prosody-typography mapping model, discussing its choice of features, typographical mappings, and implementation details. 

In Section~\ref{sec:evaluation}, we will present an experiment we ran aimed at understanding if the choices we took with the model generated typographic results that were sufficiently clear to be understandable by a general audience.

The results of said experiment are presented in Section~\ref{sec:results} and discussed in Section~\ref{sec:discussion}. Lastly, we present our broad conclusions, limitations, and proposed future work in Section~\ref{sec:conclusion}.


\section{Related work}\label{sec:related-work}

This section is divided into three main areas. We first give an overview of how prosody has both linguistic and paralinguistic functions, and the latter grounds our approach. Following, we show how different authors have created visual representations for elements of sound that are typically missing in textual transcriptions. Lastly, we present works that, like our own, have used algorithmic approaches to modulate the visual form of typography to echo sound.

\subsection{On the importance of prosody to shape meaning}

Transcribing speech into captions is a lossy process. Latin scripts are able to codify phonemes but, beyond those, the means of representing expressive variations originally present in speech are restricted to just a handful of possibilities, such as typographical emphases (variants like \emph{italic} and \textbf{bold}), special-purpose characters (!, ?, and more \cite{buchanan2015face}), emojis and creative distortions of standard grammar and the use of punctuation \cite{mcculloch2019}, etc.

While words serve to encode concepts (e.g. the word \emph{cat} can be deduced to mean the feline animal), only decoding them may not be enough to parse a statement's meaning --- or, rather, to close in on a \emph{plausible} interpretation among many possible alternatives. In a typical conversation, a listener will attempt to infer their understanding by using any available signals to close in on the most likely meaning among many credible alternatives. Prosody is one such signal \cite{wilson2006relevance}. In the \emph{a cat was standing beside that car} statement, for instance, if one's voice emphasizes the word \emph{was}, the cat might now be gone; if \emph{standing}, maybe a contrast is being made with another cat that could have been lying down; if \emph{that car}, maybe they are saying that they meant \emph{that} car and not other plausible cars? And so on. It is only upon incorporating prosody in their interpretation that a listener will be able to settle between these concurring alternative meanings, hence its importance \cite{barbosa2012conhecendo}. 



\subsubsection{Paralinguistic prosody}

One of prosody's functions is, then, to assign contrastive focus to one word in opposition to others, guiding us to sense what is more or less important at any given time. Yet, just as the voice generally has dimensions of meaning that go beyond language itself, prosody also conveys information about the speaker (e.g., age, gender, geographical origin, etc), their dispositions (are they tired, grumpy, sick, drugged, etc?), or their moods and emotions \cite{livroDoPlinio}. An~example of this varied nature of prosody, \cite{de2016prosody} showed that pitch and rhythmic information in Brazilian Portuguese was able to encode both linguistic (whether an utterance was declarative, interrogative, or imperative) and paralinguistic dimensions (e.g., categorical emotions associated with an utterance) in speech.




From a perceptual point of view, \cite{silva2016cross} show how both linguistic and acoustic interpretations of emotion in speech can be present simultaneously. In their experiment, Swedish and Brazilian participants were presented with a set of speech samples in Brazilian Portuguese, which they had to label among a set of categorical emotions. The Brazilian group had a greater agreement on the labels chosen, but the Swedish --- disconnected from the samples' linguistic but not paralinguistic dimensions --- were still able to significantly agree on the labels. 

An in depth discussion about \emph{how} speech generally \cite{koolagudi2012emotion, el2011survey}, and prosody particularly \cite{rao2010characterization}, are able to encode emotion is not the focus of this paper. Suffice to say, this is important for areas such as speech emotion recognition, a diverse field with competing approaches that consider different acoustic features (e.g., pitch, formants, energy, timing, articulation, etc) to identify emotions, themselves understood and organized by different theories and frameworks \cite{stark2021ethics}. These systems have varied applications in fields ranging from customer satisfaction evaluation to depression diagnosis \cite{perez2021user}. 

Still, given the complex nature of speech, even when using purposefully built data sets with state-of-the-art approaches, recognition rates can be generally low, both for machine-based or human classifiers \cite{ong2019modeling}. This can be seen as a testament to the complexity of the task or, alternatively, as a sign of the limits to what can be inferred when one considers acoustic signals isolated from context.


Coming from a different paradigm, \cite{Boehner2005} proposes that we see emotion not as an \emph{informational layer} embedded in speech (and therefore obtainable from the acoustic signal) but, rather, as a \emph{culturally grounded phenomenon}, echoing \cite{wilson2006relevance}'s notion that prosodic signs must be understood not in isolation, but within a given cultural and social context. 

For these authors, the same acoustic cues may have different meanings depending on their surrounding context, which imposes limits to how well an artificial intelligence system is able to perform considering only acoustic cues. While emotion is inevitably tied to the physiology of the bodies producing it, and even if it is methodologically convenient to be able to measure these physical correlates to emotion, the subjective experiencing of emotions is not constrained by only these signals \cite{boehner2007emotion}. 

Rather, we see emotion as something happening \emph{between} subjects\footnote{These can be persons, but also a person and an object, for instance.}. An example of how this theoretical framework can inform affective computing is given by \cite{angesleva2004emotemail}. In it, a software takes snapshots of a person as they are writing an email, and those will accompany each paragraph of the message on the recipient's end. The goal of this system is not to interpret the writer's emotions as they compose the message but, instead, to provide additional information which the reader can use to inform their understanding. This is similar to how we conceptualized our own model.

\subsubsection{Why represent prosody through typography?}\label{sec:why_prosody}

We base our approach not in an attempt to directly represent \emph{emotions} in speech but, rather, to represent \emph{prosody}. If we see emotion as a \emph{culturally grounded phenomenon} \cite{Boehner2005}, it follows that it carries an inherently ambiguous dimension that depends on its originating context --- who is involved (e.g., what we know and feel about someone can regulate our reactions)? What are their cultural and social backgrounds (e.g., certain groups might see some feelings as implausible in a certain setting, while for others it might seem natural)? How does their environment influence their conversation (e.g., shouting in a loud night club implies different meanings than shouting in a silent library, even if the acoustic phenomenons coincide)?

As such, it is hard to `resolve' emotions, as if there were culturally universal signs that could be inferred from physical signals carried by speech. Our approach is, then, not to try to detect emotions to then represent them but, rather, to represent the voice itself. In representing prosody, we hope to make it visible, and thus accessible for those unable to perceive sound. The viewer, we posit, will then be able to incorporate these raw acoustic elements within their given cultural and social context to inform their own understanding of speech.

\subsection{Visual depictions of speech, sound, and beyond}

Different authors have tackled the issue of limits of what can be represented by the Latin alphabet, proposing different approaches to expand to it, either through changes in the shapes and typesetting parameters of letters, auxiliary graphical elements, or both. There is evidence that readers were able to successfully incorporate some of these systems to change their reading behavior.

\cite{dosReis2011} proposed the \emph{Speechant} system of graphical elements that can be overlaid on a traditionally typeset text to give cues about its pronunciation, particularly pitch and rhythmic information, to help Portuguese speakers who were learning English as a second language. In their evaluation, students using this system were able to produce better-sounding readings of English utterances than a control group, as judged by a panel of trained phoneticists.

Similarly, \cite{waldaPhonotype} designed an extended version of the Times New Roman typeface, with special characters able to highlight and differentiate how some letters in the Dutch language can have different pronunciations depending on where they appear --- something which, they argue, hinders how non-native speakers parse written text. 

Attempting to tackle the known issue that children who read aloud in a monotonous, non-natural tone are more likely to become poorer readers later in life, \cite{bessemans2019visual} tested ways of directly representing prosody in typography. They were able to teach these visual cues to children, who were then able to use them to read aloud more expressively. 

\cite{promphan_2017} developed a typeface that was able to have its characters' shapes shift from positive valence (rounded strokes $\to$ smooth curves) to negative valence (harsh strokes $\to$ angled curves), along a continuous scale.

\cite{rashid2008dancing} worked with artists to create visual representations in closed captions of categorical emotions present in speech through animated closed captions. The recognition of these emotions was not better for the enhanced captions against a control group with traditional captions, but both hearing and \textsc{dhh} participants stated their preference for these new, enhanced captions. 

Similarly, \cite{forlizzi2003kinedit} describes the \emph{Kinedit} system, later expanded by \cite{bodine2003kinetic} to work for instant messaging, which allows inexpert users to combine multiple animation behaviors to modulate typographical attributes such as font size, color, opacity, position, rotation, etc, which can be used to express prosody, emotions, the direction of attention, characters, etc.

Aiming at more specialized audiences, some authors developed more precise, albeit likewise hermetic, rule-base visualizations of speech to aid with, for instance, prosodic analysis \cite{albert2018using, oktem2017prosograph}. These mix traditional textual transcription with graphical elements that represent acoustic features of speech such as pitch, energy, and rhythm.

\subsection{Algorithmic modulations of typographic form to represent speech}

Some authors work exclusively with typography and rule-based manipulation of typographic parameters to echo acoustic features of speech. This approach allows for applications such as speech-modulated closed captions, with potential uses-cases not only in film media but also, as we have seen, in the many scenarios where \textsc{asr}-based captioning systems have been making inroads. 

One of the challenges of such a line of work is that, until recently, typefaces' digital files mostly worked as databases of fixed shapes, one for each glyph. While one can easily change compositional parameters such as font size, position, leading, etc, the drawing of each letter was immutable --- to access a bold version of a font, for instance, you would necessarily need an additional file. This meant that, while acoustic features can be thought of as continuous, some typographic parameters would have to be changed in discrete steps. To bypass this constraint, some authors have developed custom type shaping engines, which allowed them greater visual freedom to modulate how each letter shape echoes an acoustic feature.

\subsubsection{Projects working with custom type-shaping engines}

\cite{rosenberger1999prosodic} created a system where glyphs are composed by a set of primitive shapes, each of them subject to independent manipulation: size, mapped to loudness; pitch, mapped to vertical and horizontal stretching; rhythm, mapped by the speed at which words come into the screen. Additionally, the authors approximated the Latin alphabet with a phonetic one via the use of ligatures that merged glyphs when multiple letters would represent only one sound (e.g., \emph{th}).

\cite{wolfel2015voice} created a dynamic font-shaping engine named \emph{Voice-Driven Type Design}. It allows for the modulation of a custom font's visual attributes (vertical and horizontal stroke thickness and letter width) to echo changes in prosody, which they explored for closed captions, text messaging, and expressive visualizations of poetry. This approach was later expanded in the \emph{WaveFont} system, described in \cite{schlippe2020visualizing}, which can be used in non-specialized software for the Arabic alphabet.

\subsubsection{Projects working within traditional type-setting tools}\label{sec:traditional_type_setting}

While working within the constraints of already existing type shaping engines can impose limits to how one can expressively manipulate typography, it has two important advantages: first, it allows system designers access to hundreds of thousands of typefaces and their not easily replicable features, like for example how some fonts have an extended character set to include many languages, or others have their shapes built for specific purposes (e.g. helping dyslexic readers \cite{zhu2020analysis}, working well in small, low-resolution screens \cite{lowresfont}, etc). Second, using off-the-shelf typographic engines allows for easier integration of custom typesetting algorithms into already existing workflows. 

\cite{castro2019maquinaMsc} mixed discrete (font-weight and letter repetition) and continuous (font-size and word spacing) typographic attributes to represent speech. They worked with traditional fonts, modified by scripts run in the Adobe InDesign software using the extracted prosody from audio files. As did \cite{wolfel2015voice}, both approaches were empirically evaluated, with positive recognition outcomes --- although the reliance on self-reporting of the former and exaggeratedly differentiated sound files of the latter casts some uncertainty over the results, which we hope to address in our own research.

\cite{de2020speech} ran an evaluation of what typographical attributes participants would consider the most appropriate to represent two prosodic features: loudness and pitch. They tested four typographic parameters: font weight, baseline shift, slant, and letter width. The first two were highly ranked as representations of, respectively, loudness and pitch. Of note, this approach uses variable fonts, a technology that allows for continuous variation of glyphs' shapes along pre-determined axes (we will discuss this further in section~\ref{subsec:rendering}).


\section{A model for speech-modulated \\typography}\label{sec:model}

\begin{figure*}
  \centering
  \includegraphics[width=\textwidth]{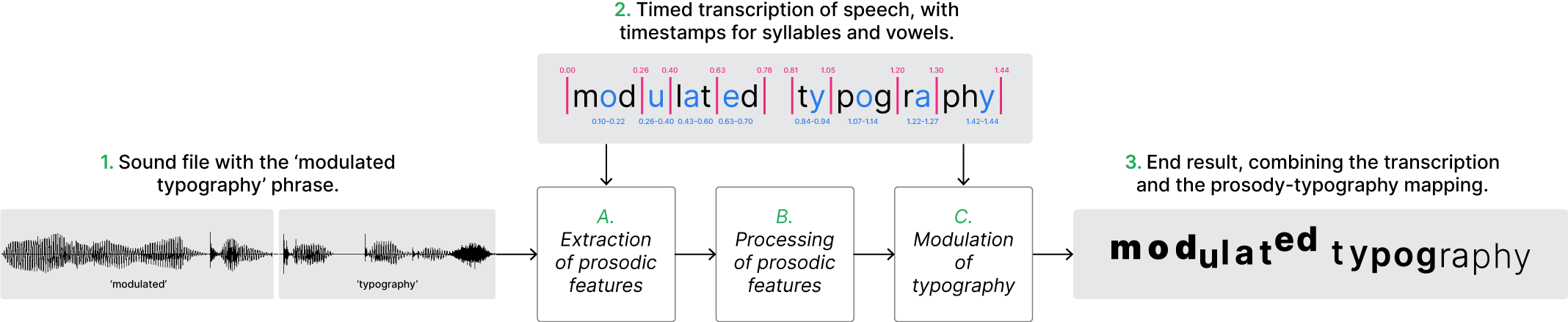}
  \caption{Diagram of how the model combines the sound of a given spoken utterance (identified by the green \imgid{1}) with its timed transcription (\imgid{2}), generating text that has visual cues for prosody (\imgid{3}). This processing is done in two parts: first, prosodic features are extracted from the audio and processed (\imgid{A} and \imgid{B}, discussed, respectively, in sections~\ref{subsec:feature_extraction} and \ref{subsec:feature_processing}). Following, these values are used to modulate typographic parameters (\imgid{C},~discussed in section~\ref{subsec:modulation_of_typography}).}
  \label{fig:model_diagram}
\end{figure*}

In this section, we will present a model that extracts meaningful acoustic features from speech and, after they are processed, uses them to modulate parameters of typographic shape and composition. An overview of this process can be seen in figure~\ref{fig:model_diagram}. 

\subsection{Prosodic features}

\subsubsection{Extraction of prosodic features}\label{subsec:feature_extraction}

To represent speech in typography, we modeled a set of acoustic dimensions that are related to how the voice's expressiveness changes over time. We worked with three acoustic measures: \emph{magnitude}, \emph{pitch}, and \emph{duration}. 

One of the advantages of representing prosody and not other acoustic features, despite the good results in emotion recognition systems obtained with other metrics (e.g., \cite{eyben2015geneva}), is that with prosody the mapping between acoustic feature and written units is direct. By extracting features considering the unit of one syllable, we obtain values that can be then meaningfully mapped to this same syllable's textual representation. 

In figure~\ref{fig:model_diagram}, this process is represented by \imgid{A}, which receives as inputs the sound file \imgid{1}, from which the features are extracted, and the timed transcription \imgid{2}, which defines the timestamps to subdivide \imgid{1} into each of its syllables.

The first feature is related to perceived loudness. It is calculated by the root mean square (\textsc{rms}) of all samples in a given region\footnote{While the two other prosodic features were calculated using audio segments that corresponded to whole syllables, for magnitude we only considered excerpts that matched vowels, excluding consonants.}. Given an array with $k$ audio samples $A = [a_1, a_2, ..., a_k]$, \textsc{rms}~is calculated by:

\[
A_\textrm{rms} = \sqrt{\frac{1}{k} \sum_{j=1}^{k} a^2_j}
\]

\emph{Pitch} is related to the perceived melody of the voice and was calculated through an auto-correlation algorithm, as described by \cite{boersma1993accurate}, applied at the level of the syllable, limited to between 50 and 350 Hz (typical for a masculine voice, as was the case for the dataset used in our evaluation). 

Lastly, \emph{duration} relates to rhythm and is the simple temporal measurement of each syllable.

\subsubsection{Processing of prosodic features}\label{subsec:feature_processing}

While magnitude, pitch and duration are measured absolutely, they are perceived relatively. This means that each syllable is perceived not by its absolute values, but rather by how it is related to its neighboring syllables. Prosody creates contrast: \emph{this} is different than \emph{that} \cite{barbosa2012conhecendo}. 

Because of this, to modulate typography we must use not the original prosodic measurements, but rather their relative values. In figure~\ref{fig:model_diagram}, this transformation of the original acoustic features extracted in \imgid{A} is represented by \imgid{B}. These values are obtained by normalizing each prosodic measurement considering maximum and minimum values at both an utterance and local level. For the former, we used:
 
 \[
x_i' = \frac{x_i - x_{i\textrm{ min}}'}{x_{i\textrm{ max}}' - x_{i\textrm{ min}}'}
\]

\noindent where $x_i$ is the unprocessed obtained value for each feature, and where $x_{i\textrm{ min}}'$ and $x_{i\textrm{ max}}'$ are, respectively, the minimum and maximum values for that same feature considering the whole utterance. To calculate the \emph{local} normalization, we used:

\[
x_i'' = \frac{x_i - x_{i\textrm{ min}}''}{x_{i\textrm{ max}}'' - x_{i\textrm{ min}}''}
\]

\noindent where $x_i''$ is a normalized value for the $x_i$ feature that considers an unequally-spaced window of 15~syllables around it, as such:

\[
x_{i\textrm{ max}}'' = \textrm{ max}\{x_j\}^{j=i+5}_{j=i-10} \textrm{ and } x_{i\textrm{ min}}'' = \textrm{ min}\{x_j\}^{j=i+5}_{j=i-10}
\]

\begin{figure}[h]
  \centering
  \includegraphics[width=.9\linewidth]{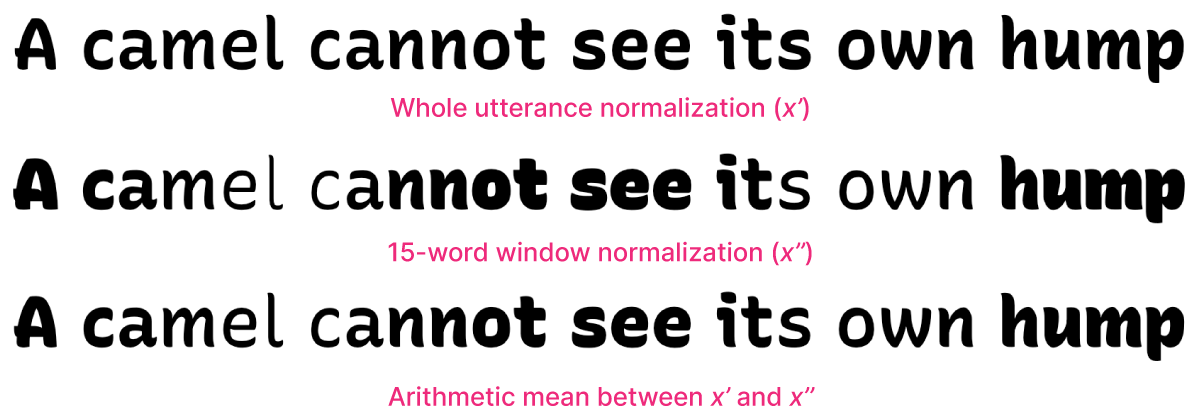}
  \caption{The three normalizations. From top to bottom, the one that considers the whole utterance, the one that considers a window of 15 positions around the current one, and the arithmetic mean of the two. To accentuate the effect for illustrative purposes, in this figure we have used random numbers applied to each \emph{letter} --- rather than to each syllable, as was used in our model.}
  \label{fig:camel}
\end{figure}

The final value was an arithmetic mean of $x_i'$ and $x_i''$. An example of how the two variables produce different normalizations is seen in figure~\ref{fig:camel}.

\subsection{Modulation of typography} \label{subsec:modulation_of_typography}

With the prosodic features processed and ready for use, the next step was to map them as transformations of typographic parameters. In figure~\ref{fig:model_diagram}, these features come from \imgid{B} (processing of prosodic features) and, with the transcription coming from \imgid{2}, are processed by \imgid{C} (modulation of typography) to generate the end-result in~\imgid{3}. 

As mentioned in Section~\ref{sec:traditional_type_setting}, in \cite{de2020speech} participants ranked how well different visual parameters were able to represent the prosodic features of magnitude and pitch, for which a clear preference arose for the use of, respectively, font-weight and baseline shift. We followed this recommendation in our current work.

\begin{figure}
  \centering
  \includegraphics[width=.9\linewidth]{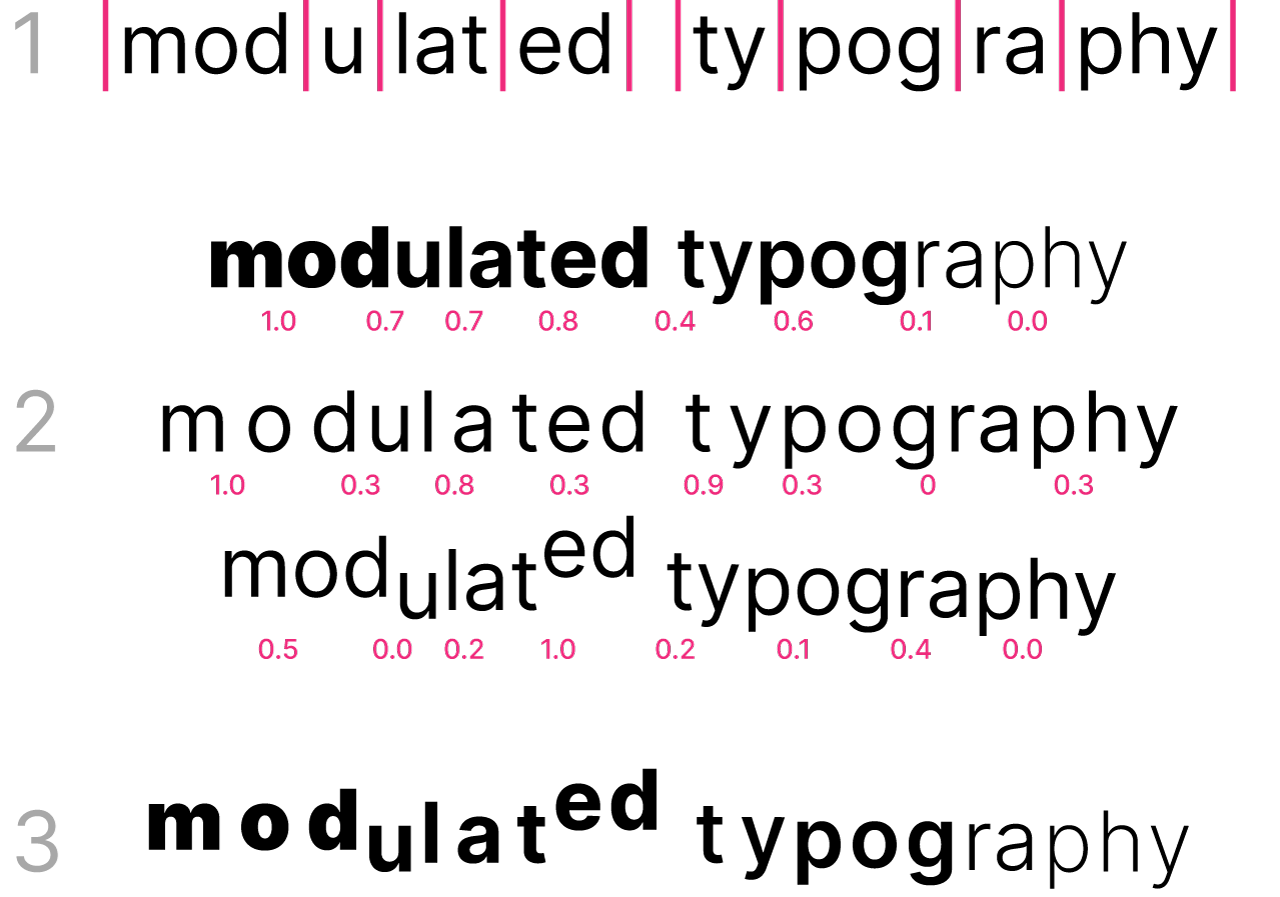}
  \caption{The words `modulated typography' are shown with a decomposition of how the model will modulate each syllable using the three typographic parameters used in our model. On the row identified by the number 1, the words are in their default state with the syllabic boundaries highlighted (for this example we are using the \emph{Inter} typeface). On row 2, each of the three typographic modulations is represented, one in each line. They are, respectively, font-weight, letter-spacing, and baseline shift. Underneath each syllable, in pink, we marked the relative values each parameter received. On row 3, the finished product, with the three typographic modulations applied at the same time.}
  \label{fig:modulated_typography}
\end{figure}

\emph{Font-weight} is a parameter that sets the thickness of each letter, e.g., higher values make letters thicker, while \emph{baseline shift} controls the vertical displacement of letters. While these changes can be discrete (e.g., most fonts will have separate files for a regular and bold version), we favored variable fonts with an axis for font-weight, allowing for continuous changes. In the example shown in figure~\ref{fig:modulated_typography}, we used the \emph{Inter} typeface; in figures~\ref{fig:camel} and \ref{fig:screenshot} (and in the evaluation itself), we used the \emph{Recursive} typeface. Both are available on the Google Fonts website \cite{andersson_nixon}.

\emph{Baseline shift} is a compositional parameter. This means that it is not related to the shape glyphs themselves but, rather, to how they are placed on the line. As such, it is independent of the typeface used.\footnote{It is worth mentioning, however, that fonts with long ascenders and descenders (the parts of the letters that extend beyond its main body, such as the tail in a `j' or the hook in an `f') will allow for smaller variations in baseline shift before lines start overlapping.}

We were still left with having to find a typographic modulation to represent the duration of syllables. While \cite{wolfel2015voice} and \cite{bessemans2019visual} explored using letter width to echo speech's rhythmic patterns, \cite{de2020speech} points that it, along with slant, are not ideal modulation candidates --- they were rarely chosen over font-weight and baseline shift and, when they were, they served more as a way to represent a voice that was flat, i.e., inexpressive, than prosodic variations per se. 

Inspired by how \cite{castro2019maquina} represented the duration of pauses through changes in spacing between words, we used the compositional parameter of \emph{letter-spacing} (the horizontal spaces between each letter) to represent each syllable's duration. Following \cite{chung2002effect}, who shows that decreasing letter-spacing can reduce reading speed, we worked only with positive values, avoiding \emph{squeezing} letters together. Thus, a faster-than-average syllable was displayed with normal letter-spacing, while a slower than average one would have wider spacing.

The complete model, with the three prosodic-typographic mappings, can be seen in action in figure~\ref{fig:modulated_typography}. From the image, we can deduce that the \emph{modulated} was spoken at a louder volume than \emph{typography}, and that its \emph{mod} syllable was the slowest and that its \emph{ed} syllable the one with the highest pitch. Some intuition of these patterns can be seen in the sound wave representation seen in figure~\ref{fig:model_diagram}, which is from the recording used to generate this example.


\subsubsection{Speech-modulated typography and variable fonts}\label{subsec:rendering}

\begin{figure}[h]
  \centering
  \includegraphics[width=.5\linewidth]{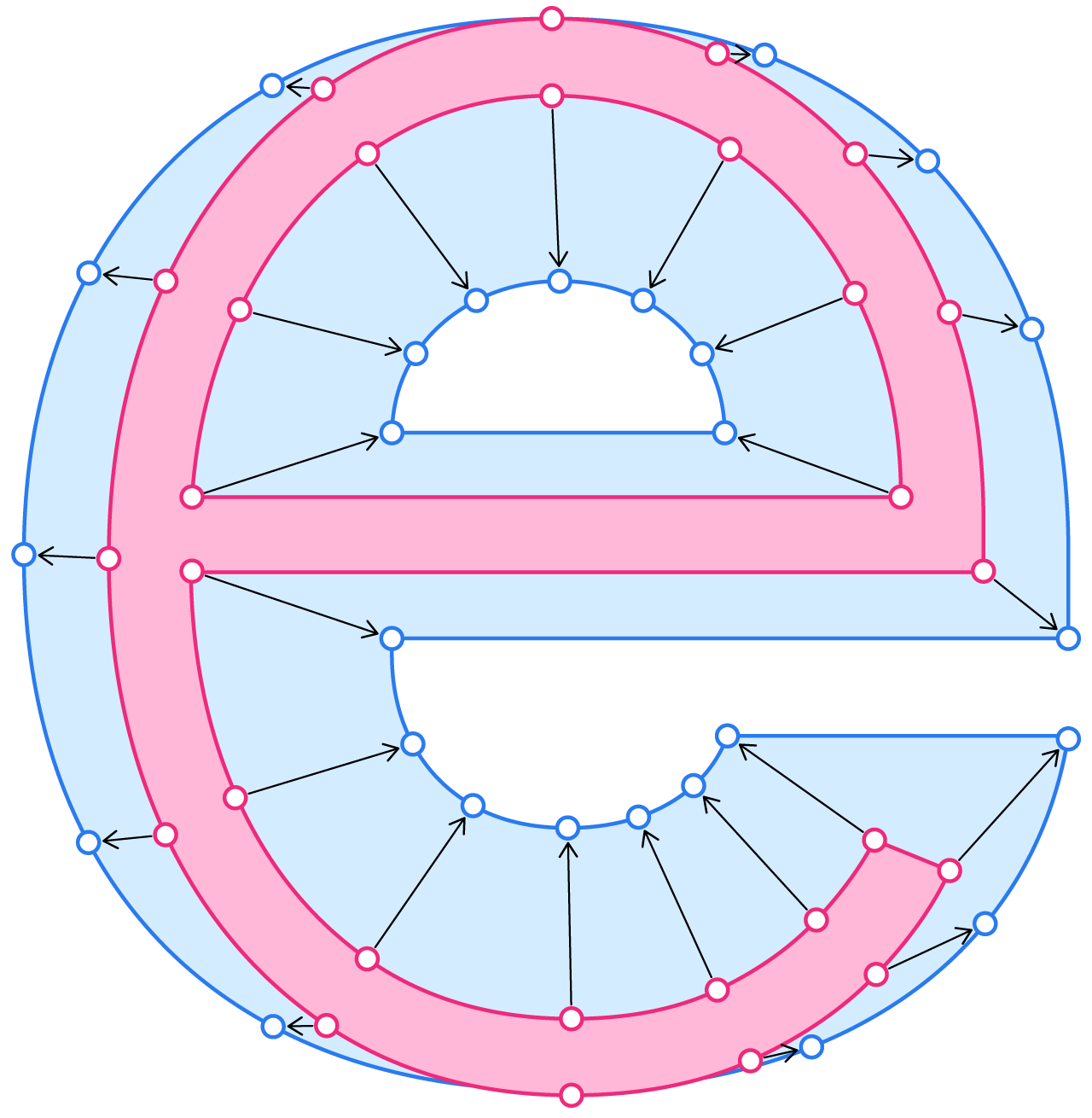}
  \caption{Example of how a single glyph changes shape between two extremes of an axis of variation. In this case, we are showing how the font-weight axis changes the letter \emph{e} from the Inter typeface, going from the lighter value of 200 (in pink) to the bolder value of 900 (in blue).}
  \label{fig:decomposed_glyph}
\end{figure}

To actually render the modulated typography, we decided to work with variable fonts because of the flexibility it allowed. Introduced in 2016 OpenType's 1.8 specification, variable fonts are a format where each glyph's shape is defined not as a collection of points with fixed $x$ and $y$ positions but, rather, by shifts in these $x$ and $p$ positions along \emph{axes of variation} \cite{openType18}. An example of this process is shown in figure~\ref{fig:decomposed_glyph}. Note how some points travel long distances while others remain static --- a change in the axis' value can mean a big shift in position for some points, but little to none for others.

While we only used one variation axis in our model (font-weight), variable fonts allow for an arbitrary number of axes to be defined. Since each axis defines the changes in position for each point (a set of deltas) when more than one axis is present their combined values can be calculated by summing these differences, which allows for independent control of each axis. In the future this could help expand and modify the currently proposed model.
\section{Evaluation}\label{sec:evaluation}

\begin{figure}[h]
  \centering
  \includegraphics[width=\linewidth]{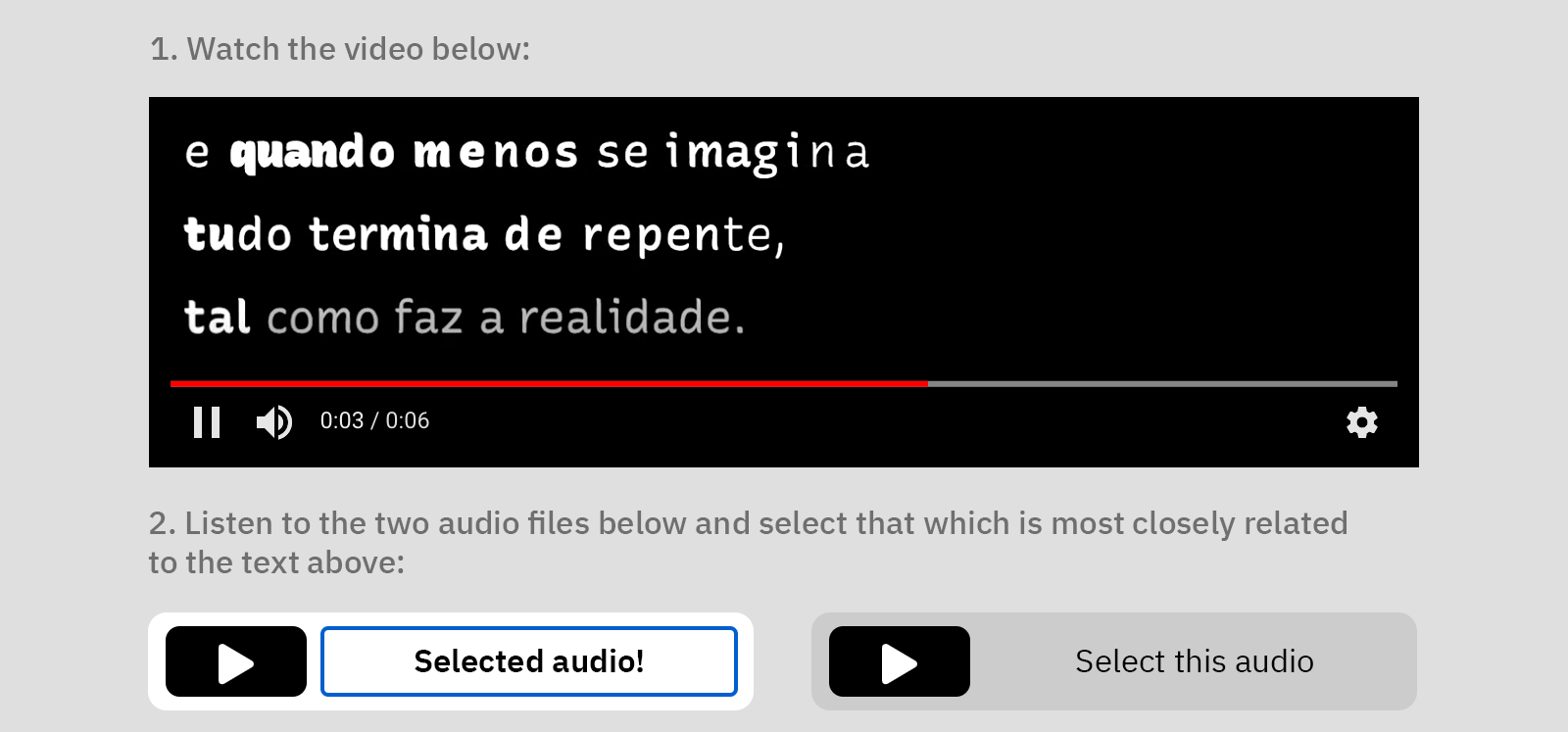}
  \caption{A translated screenshot of one round of the test. Participants had to match one of the audio files below with the video above, which was muted. Both audio files had a reading of the same text as presented in the video, but with different prosodic emphases. The words on the video say, in Portuguese, `and when one least expects it / everything ends suddenly, / as does reality.' \cite[p.~75]{britto2003macau}}
  \label{fig:screenshot}
\end{figure}

Our evaluation consisted of using our model to create the visual representations of specific speech utterances and then showing participants these representations along with two different audio files, one the original recording, another with the same text spoken with different prosodic patterns. The goal was to measure how successful participants would be in matching the speech-modulated typography to its originating audio, which was only subtly different from the other audio it was presented alongside with.

We proposed this experimental method to evaluate whether our model allowed for the \emph{recognition of vocal prosody}: a greater-than-chance rate of success for this matching between the typography and its corresponding, correct audio, would indicate that the model's visual representation of prosody was sufficiently clear to allow participants to decode its correspondence with audio. (The~evaluation was reviewed and approved by an \textsc{irb}.)

\subsection{Expressive prosody dataset}
The sound recordings used in the test were created for the purposes of our evaluation. Like \cite{castro2019maquinaMsc}, we decided to base our evaluation on readings of poetry, since they can accommodate a greater variation in vocal expression without sounding \emph{artificial} --- even if the poem's author had a specific set of emphases in mind when writing a poem, readers are free to have their own different interpretations, a freedom we took advantage of. However, because we need two different readings for each poetic stanza, we realized we would need to create our own dataset.

We instructed a hired actor to repeatedly record a reading of the same poem, at each round emphasizing different aspects of their voice: loudness, pitch, rhythm, a mix of the three, a monotonic reading, a naturalistic reading. With these different versions, we sliced the audio files into word-sized units, which we recombined into the two final versions of each poem, making sure that the three prosodic features had heterogeneous patterns between the readings.\footnote{A pilot version of our experiment showed us that these versions had to have roughly the same duration otherwise participants would try to sync video and audio to find the correct match, an effect that \cite{rosenberger1998prosodic} had already found in her own similar experiment.}

\subsection{Rendering speech-modulated typography}

To display the modulated typography generated after processing the audio files, we created a script that ran in the browser, receiving as input the video and a subtitle file that, beyond the transcription itself, contained cues about how to change the typography of each syllable, how long these changes should take (i.e., the duration of the syllable), and at which moment (i.e., the moment that syllable was said). 

Since variable fonts are already supported in all major browsers \cite{varfontbrowsersupport}, by working in a web environment we were able to take advantage of certain built-in facilities, such as how a native \textsc{html} video player can fire syncing events related to when each subtitle text block is supposed to enter or exit the screen once the video is played, or how \textsc{css} allows for typographic parameters to change with an eased, non-jarring animation. However, while we were able to run the animations in real-time in our own setups, to accommodate for participants in lower-end systems such as older mobile phones, we decided to screen capture the animations and make them available as YouTube videos.

\subsection{Test platform}

In the experiment, participants would be randomly assigned to match either animated of static instances of speech-modulated typography. The test itself consisted of 15 rounds of sound-and-image matching. Participants would be given no instructions about the prosodic-typographic mappings used by our model, which they would have to figure out on their own. Figure~\ref{fig:screenshot} shows a screenshot of the test in action.

We divided our participants into two groups: the first received a \emph{static-image} based version of the test, i.e., each of the 15 stanzas was displayed as an image to which all typographic modulations were already applied. This is similar, for instance, to how the typography is shown in the third row of figure~\ref{fig:modulated_typography}.

\begin{figure*}
  \centering
  \includegraphics[width=\textwidth]{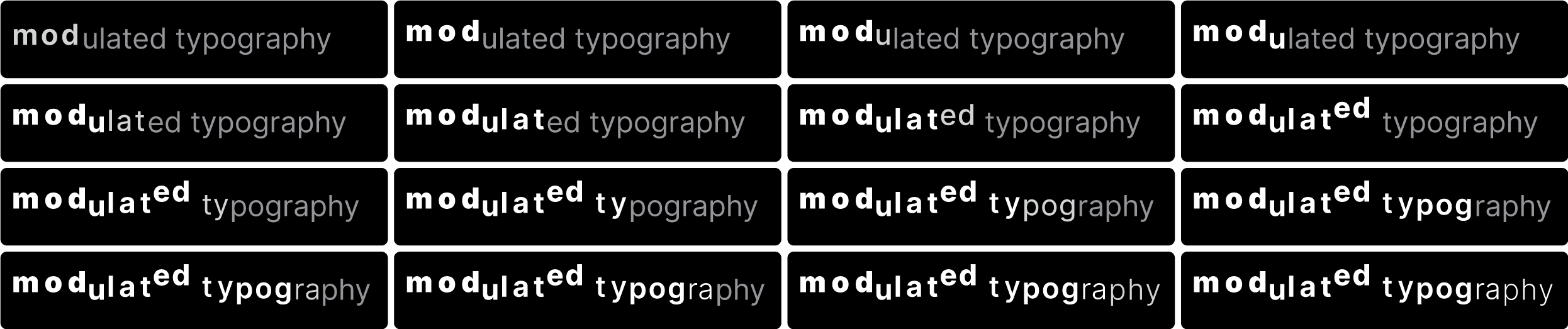}
  \caption{Example of how each syllable's typographic parameters change in time, as was shown to participants that did the animated closed-caption based test. The image is meant to be read from left to right and, subsequently, from top-to-bottom.}
  \label{fig:frame-by-frame}
\end{figure*}

The second group of participants did an animated closed-caption version of the test. This consisted of the same stanzas and audio files, but each version of speech-modulated typography was provided as an embedded, muted YouTube video. This video started with a version of the stanza in its default state (i.e., neutral typographic parameters), and as each syllable sounded its typographic parameters animated towards their end state (see figure~\ref{fig:frame-by-frame} for a frame-by-frame example). 

By showing changes synchronously with audio, these animations emphasize the rhythmic aspect of the model, but it can also make it noisier to parse, possibly adding a cognitive cost for readers. Having both this animated and the static version of the test running in parallel aimed to answer our second research question.

\subsection{Open-ended comments}

At the end of the evaluation, participants were prompted to leave us comments about `the test, speech-modulated typography, its possible applications, or whatever else they wanted.' This step was optional, but nevertheless 43 participants took it. We analyzed these messages using the thematic analysis approach, outlined in \cite{braun2006using}: after familiarizing ourselves with the messages, we generated an initial batch of codes, which were then reviewed and collated having our research questions in mind, but also our additional goal of helping to uncover limitations of our model and methodological approach to evaluate it. Lastly, these codes were summarized into two themes, which we will present in section~\ref{sec:results_comments}, and discuss in section~\ref{sec:themes_discussion}.

\section{Results}\label{sec:results}

\subsection{Participants}
We ran the experiment from November to December 2020. Participants were recruited through social media and academic mailing lists. Of the 117 participants that concluded the test, 47\% were female, 51\% male, and 2\% non-binary; 90\% had undergraduate degrees or higher; 25\% had between 18 and 24 years, 49\% between 25 and 39, 23\% between 40 and 59, and 3\% had 60 or more years. 52\% of the participants were randomly assigned the static-image-based test, while 48\% received the animated closed captions variant.

\subsection{Rates of success}

\begin{figure}[h]
  \centering
  \includegraphics[width=\linewidth]{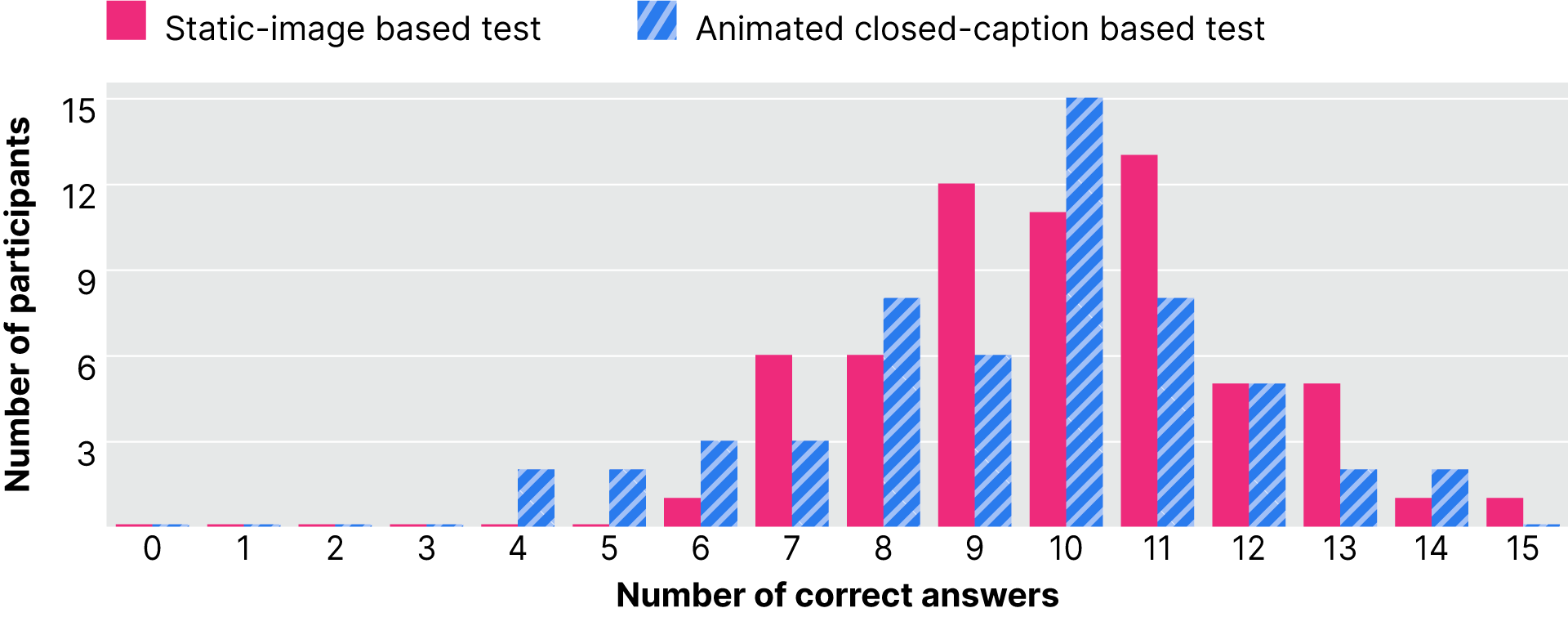}
  \caption{Distribution of how many participants (in the y axis) got how many correct answers (in the x axis), considering  both versions of the test.}
  \label{fig:results_graph}
\end{figure}

The average rate of correct answers for the 61 participants who took the static-image-based test was 67\% (95\% \textsc{ci} [.64, .70]), slightly higher than the 63\% (95\% \textsc{ci} [.59, .67]) obtained by the 56 participants who took the animated closed captions variant, as shown in figure~\ref{fig:results_graph}. This 4\% difference between the two distributions was found to be not significant ($t(115)=1.54, p>0.05$).

\subsubsection{Effect size}

To simulate a control group we assumed that an ineffective prosodic-typographic model would have no effect on participants, generating results indistinguishable to those of a random distribution. We compared this simulated data to our actual results to measure the effect size of each of our two approaches\footnote{Results reported averaged over 1,000 runs of the statistical tests, after which fluctuations of the randomly generated values stabilized.}. Both the static-image-based test ($t(59)=4.18, p<0.05, d=0.75$) and the animated closed caption-based variant ($t(54)=2.91, p<0.05, d=0.55$) were significantly different from the random distribution, with effect sizes of \emph{medium} magnitudes (Cohen's d between 0.5 and 0.8).

\subsection{Open-ended comments}\label{sec:results_comments}

Sending us comments at the end of the evaluation was optional. Nevertheless, 43 out of the 117 participants left their thoughts. In the quotes below, participants will be marked by the letter \textsc{p} along with an identifying number. We have translated the comments, which were originally in Portuguese, and, when necessary, edited for clarity. We summarized the data using two themes, which name the following sections.

\subsubsection{Theme 1: Divergent interpretations}

Regardless of how well they did towards correctly matching audio and typography, we received comments both complimenting how easy the model was to understand or criticizing how obscure it was. Among the discontents, \p{1} commented that baseline shift was not an intuitive modulation and that some more subtle differences in intonation became ambiguous, which echoed \p{2}, that among all modulations found those the hardest to associate with the audio. \p{3} tried, but could not find a clear logic behind the modulations, something also present in \p{4}'s comment that `each round seemed to have a unique pattern,' and in \p{5}'s that `the meaning behind how letters moved wasn't clear.'

Some participants found the model easier to grasp. \p{6} understood that thicker letters represented louder sounds, while \p{7}, \p{8}, and \p{9} all managed to correctly match the three typographic modulations (font-weight, baseline shift, letter-spacing) to their corresponding prosodic features (respectively, magnitude, pitch, and duration). \p{9}, particularly, said that:

\begin{quote}
    When seeing the video without the audio and then comparing the two [audio] options it was perfectly possible to interpret, using only the sound, when the actor paused, elongated words, shouted, made their voice thicker or thinner. It was not possible, however, to extract from the text whether his voice was calm, commanding, irritated, etc.
\end{quote}

\subsubsection{Theme 2: Expansions to the model}

Some participants felt that the model should be changed and/or expanded to include additional typographic modulations. \p{10} suggested that we could have tested also using uppercase words to represent emphases. \p{1} suggested that we should have explored a visual language closer to that of comic books. Similarly, \p{7} thought that to be able to represent voice qualities such as a raspy voice, the fonts should have pointy corners, while a soft voice could have been represented by smoother corners (similar, then, to \cite{promphan_2017}'s proposed font). They go on to say that

\begin{quote}
    [it should be] more cartoonish, more expressive. How would these captions represent different voices that are feminine, elderly, young, stuttery, twangy, sleepy, non-native, [etc]?
\end{quote}

There were comments that made reference to how our experiment's format made the text hard to understand. \p{11} told us that at each step they felt the need to know how the actor had read aloud the previous stanzas, since understanding `if a clause is subordinate or coordinate (...) has great impact in how the voice is modulated.'

We received some comments about the model being hard to read. \p{5}, for instance, suggested we `make the text bigger and more legible.' \p{12} mentioned being dyslexic, and that they had a hard time reading the texts, 
\begin{quote}
    since the movement of the letters and changes to their typography disrupted me, and I had to watch the videos many times before I was able to understand them.
\end{quote}

\p{13} wrote that, when words had their letter-spacing increased, it became hard to differentiate one word from the other:

\begin{quote}

For example: at first glance, I considered that `\textsc{a~d~o~r}\footnote{This is not a word in Portuguese.}' was a single word, waiting for the rest [that would complete the word], but eventually I realized that there were two words, `\textsc{a dor}\footnote{`The pain,' in Portuguese.}.' I had to read the captions again to make sense of them.
    
\end{quote}

\section{Discussion}\label{sec:discussion}

We asked whether \emph{visual modifications in the shapes of letters composing a text could allow for the recognition of vocal prosody.} Our experiment was able to measure a strong-enough effect to affirm that yes: participants were able to intuitively grasp our prosodic-typographic mappings in such a way as to be able to \emph{reverse-engineer} the images and find their originating audio files. This reinforces \cite{de2020speech}'s proposed mapping of font-weight and baseline shift to, respectively, magnitude and pitch, as it does to our addition of letter-spacing mapping changes in duration.

Curiously, and in answering our questions of whether there would be \emph{differences in recognition between animated and static versions of speech-modulated typography}, the animated typography did not seem to make a difference in terms of increasing (or decreasing) recognition performance. While this is an indication that static-image-based speech-modulated typography could find useful applications on its own, it is also surprising considering that in the animated tests there were additional signals helping tie images to sound.

\subsection{What did participants tell us?}\label{sec:themes_discussion}

While our quantitative results point to our model's success in conveying speech through typography, participants' answers paint a more nuanced picture. Many of them were able to correctly identify each of the model's mappings between prosody and typography, while others found it at times inconsistent, with some specifically pointing out problems with our use of baseline shift. These divergent interpretations show that, even if the model as a whole was effective, more work is needed to understand how well each typographic parameters is able to represent its corresponding prosodic feature, and whether there are perceptual and semantic changes when they are used in tandem.


Moreover, and perhaps surprisingly, some participants felt the model should be able to embody more nuanced and complex representations of speech, which it should do with a more varied and expressive visual vocabulary. While this is not a simple task, it is nevertheless encouraging to imagine how different applications of speech-modulated typography could branch out as this new field develops.

Care must be taken, however, to whether the modulations can hinder legibility. The use of animations, and the change of positioning and spacing parameters caused discomfort to some participants. Future work is needed to understand how models such as the one we propose here could make text harder to read, and applications of these models should consider giving the user options to either tone down the modulations or turn them off completely.


\section{Conclusion}\label{sec:conclusion}

We have proposed a model for transforming acoustic cues present in speech into visual modulations of typography, allowing for a transcription of not only words but also the paralinguistic dimensions of an utterance. This model was evaluated with hearing individuals that, having had no previous training, were able to use this speech-modulated typography to correctly match speech and typography. The model --- which mapped magnitude, pitch, and rhythm to, respectively, font-weight, baseline shift, and letter-spacing --- worked equally well when its modulations were presented as animations or static images.


Although we envision that our approach could be beneficial to \textsc{dhh} individuals, the fact that we conducted an evaluation with only hearing participants limits how generalizable our results are, particularly considering \textsc{dhh} persons --- it is plausible, for instance, that a firsthand based intuition of speech might be needed for one to make sense of our model.


Future experiments should attempt to investigate how well the model performs with \textsc{dhh} persons, and what adjustments could me made to it considering their perceptions of speech-modulated typography. Considering how \textsc{asr}-based captioning systems have made in-roads in many contexts, understanding if and how our approach could bring benefits to specific settings could also be fruitful.

The claims by some participants about how the model is at times difficult to read should be investigated.

Lastly, and in addition to simply measuring the model's \emph{recognition} rate, an important avenue for future exploration should also include attempting to understand what \emph{effects} these speech-modulated closed captions could have on viewers' subjective experience, e.g., immersion in film-media, quality of communication in online meetings.

\ifCLASSOPTIONcaptionsoff
  \newpage
\fi



\bibliographystyle{IEEEtran}
\bibliography{IEEEabrv,./references}
%



%


\begin{IEEEbiography}
    [{\includegraphics[width=1in,height=1.25in,clip,keepaspectratio]{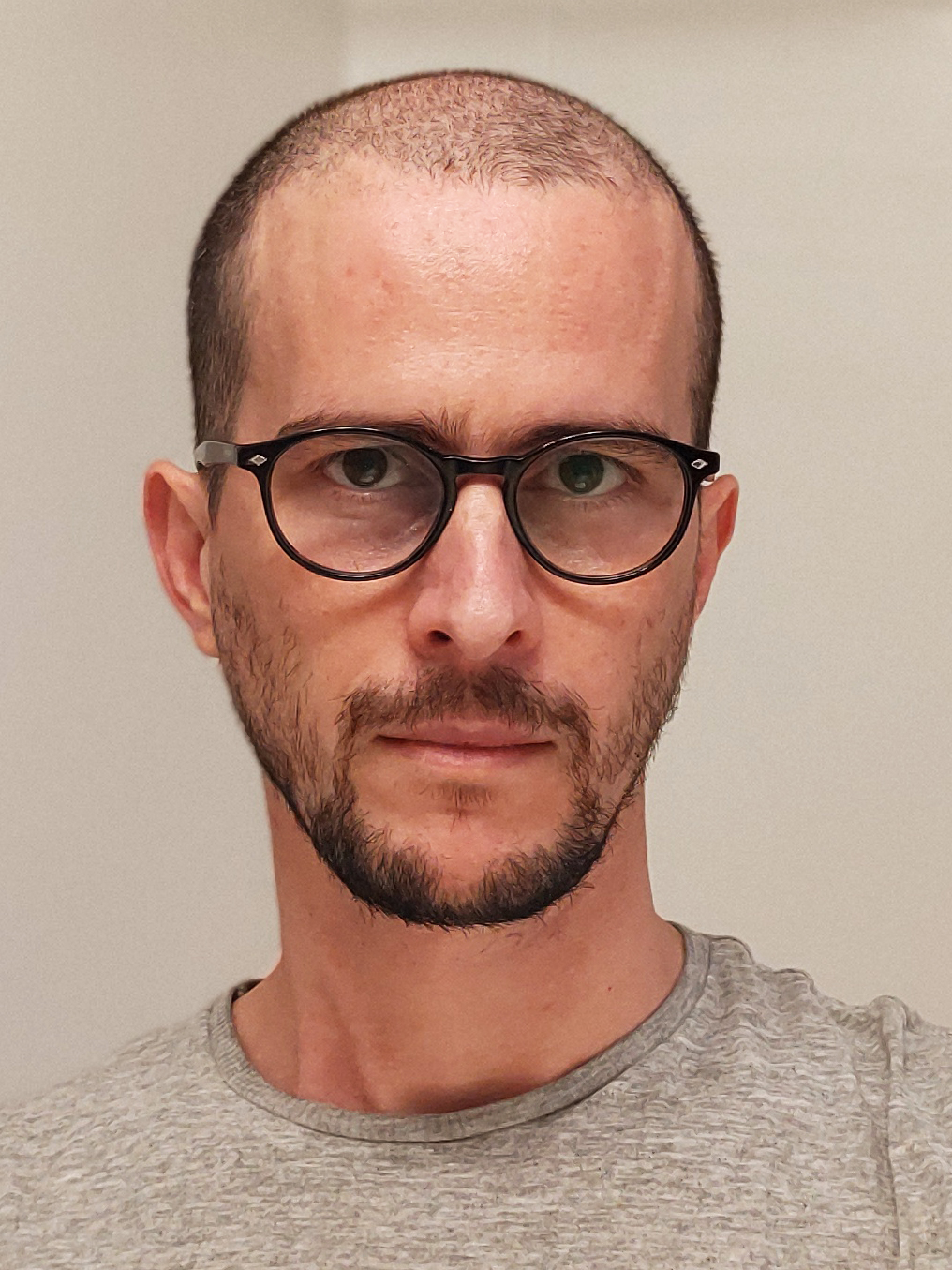}}]{Caluã de Lacerda Pataca}
received his MSc degree in Computer Engineering from the University of Campinas, Brazil, in 2021, and is currently working towards a Ph.D. degree at the Computing and Information Sciences department at the Rochester Institute of Technology, USA. His research interests include speech accessibility, visual design, and human-computer interaction.
\end{IEEEbiography}

\begin{IEEEbiography}
    [{\includegraphics[width=1in,height=1.25in,clip,keepaspectratio]{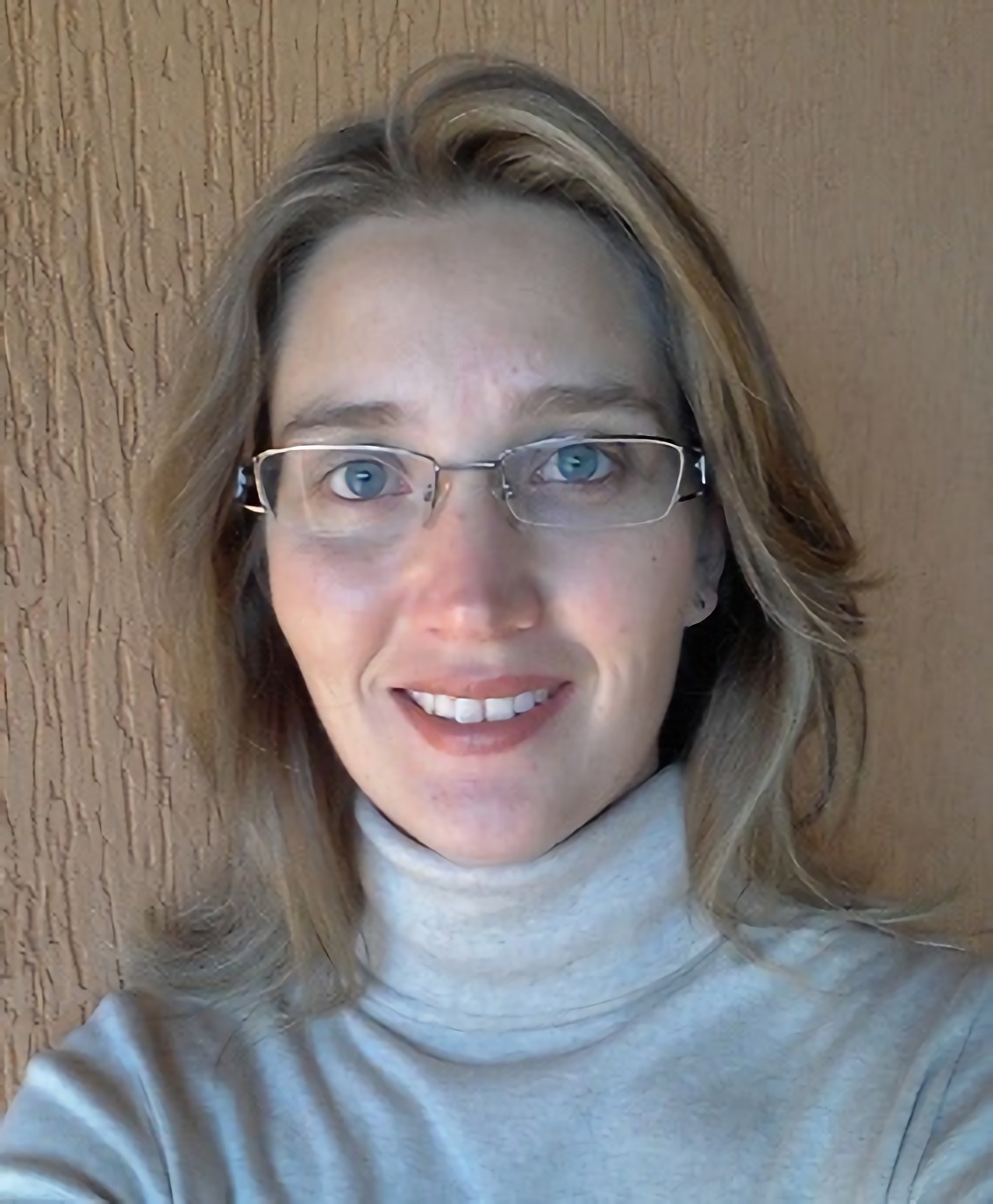}}]{Paula Dornhofer Paro Costa} received her Ph.D. degree in Computer Engineering from the University of Campinas (Unicamp), Brazil, in 2015. In 2016, she joined the Dept. of Computer Engineering and Automation (DCA) of the School of Electrical and Computer Engineering, at Unicamp, as a research scientist and assistant professor. Her research interests focus on affective computing and multimodal artificial intelligence.
\end{IEEEbiography}

\vfill







\end{document}